\documentstyle[aps,epsf,prl,twocolumn]{revtex}
\begin{document}
\twocolumn[{
\draft
      
\title{Short,  Medium and Long Range Spatial Correlations in Simple Glasses}

\author {Tamar Kustanovich  and  Zeev Olami}
\address{Department of~~Chemical Physics,\\
 The Weizmann Institute of Science,
Rehovot 76100, Israel}
\maketitle
\widetext
\begin{abstract}
\leftskip 54.8pt
\rightskip 54.8pt
%%%%%%%%%%%%%%%%%
Local stresses and pressures  always exist in glasses.
 In this letter we consider their 
effects on the structure and structural  correlations in simple glasses. 
We find that extreme values of local pressures
are related to well defined local structures. The  correlations related to these
 extreme stresses 
extend to full system size and
decay as a power law with the distance.
 This result is especially striking, since at large scales, the total 
density fluctuation exhibits 
 exponentially damped decay similar to the decay in simple liquids. 
Thus at medium and  large distances,
 the atoms  with extreme values of local pressures exhibit higher degree of 
correlation than the rest of
 the system.   These results 
 were found for glasses with very different short range structure, indicating 
 their general nature. 
 %Thus, the arising picture of glass is as stressed network
 %that induce short, medium and long range structural correlating.   
\end{abstract}

\leftskip 54.8pt
}]
\narrowtext
The nature of correlations in amorphous solids and super cooled liquids poses a
fascinating question that has  received a lot of  attention (see e.g. \cite{eli,Don}).
While in crystals there is a clear cut definition of structural correlations
that uses the period of the crystalline lattice, how to define structural correlation
in glasses and liquids is an open question that may have various answers.

A related question is which physical quantity should be used in looking
for correlations.  There are various possibilities.
 The simplest one is to analyze the average 
number of atoms in a distance
$r$ from a  central atom, using the radial distribution function (RDF).
 For simple liquids with a known atomic pair potential, the RDF
 can be calculated 
 from integral equations  supplemented by a closure 
relation and requirements for  thermodynamic consistency \cite{Han}. The
 same methods cannot be applied for the glassy phase,
 because glasses are  non ergodic systems.  
The RDFs describe average radial correlations and
thus fail to touch the finer details of the structure for glasses 
and liquids. They tend to average out variances in the 
local structure. 

A second option is to find some strongly varying
 local property of the system, like the local stress, distribution of energy 
wells, Voronoi volume or another local structural parameter, 
and use it in the calculation of conditional correlation
functions.    An obvious question in such case is whether
these parameters are manifestations of random statistical noise 
or do they exhibit a more basic nature of the glass.

Local stresses and pressures always exist in glasses. 
They result from  variations in the local environments,
which persist even in a zero-temperature glass, where all the 
local forces on the 
atoms vanish. The average stress and pressure might be zero, 
but the local stresses have a wide distribution
\cite{egami,Alexander,Kst,Kust1}. 
The glass can be described as  a discrete atomic  network  in which 
the stresses
balance each other throughout the network.
 To understand such a network one should study
 the correlations  between the stressed configurations.
The role of  the most compressed/stretched sites is an important issue.
We  show that they are manifestation 
of the basic structure of glasses.
 Thus they are relevant for classification of
 the local structural units and the degree of randomness in a glass. 
 
 The characterization  of stressed  networks is naturally complicated since they 
involve tensorial description and statistical noise. 
However, even without knowledge of the full details of the network, 
one can  use the local stresses to obtain an elastic and structural
 description of a glass \cite{egami,Alexander,Kst,Kust1}. 
  An appropriate scalar, such as the trace of the site stress 
 tensor\cite{egami,Alexander,Kst}
 or the local pressure\cite{Kst}, can be used as a parameter for the study of 
 correlations in the glass. In this letter we present our study of the effects of extreme values
 of local pressures on structural correlations in simple glasses.

For mono-atomic systems where atoms interact through a pair potential
 $U(r_{ij})$,
 where $r_{ij}$ is
the distance between atoms $i$ and $j$, we define the term \cite{Kst}
\begin{equation}
J_i=\sum_j \frac{d U(r_{ij})}{dr}
\end{equation}
as "local effective pressures" (LEP). 
  $J$'s so defined are related to an effective pressure
 $p_{i}^{eff}$ on a sphere of radius $r$ around atom $i$ by $p_{i}^{eff}=J_i/(4 \pi r^2)$. 
 Since the local pressures depend on the choice of the cell \cite{Kust1}, 
 using $J$'s as the conditional parameter enables us to
 investigate directly the effect of simple elastic features
 on structural correlations. Unlike 
 the forces $\vec{f}_i=\sum_j \vec{ \nabla} U(r_{ij})$ that
 vanish at zero  temperature, LEP as well as local pressures do
 not vanish in amorphous solids,
 even when the total (bulk) pressure is zero. 
The probability distribution of $J$ is generally quite
wide, although most of the atoms feel relatively small pressures
(see examples in \cite{Kst}).

The conditional radial pair distribution function $n_J(r)$ is
defined as the average number of
neighbors at distance $r$ for atoms with a given $J$ \cite{Kst}.
The extra structural correlations, induced by stresses, can be calculated from the
function  
 \begin{equation} \label{dg}
dn_J(r) = n_J(r) -n(r)
\end{equation}
where $n(r)$ is the the average number of neighbors at distance $r$.
This  provides a measure of the difference between the stressed sets and the 
total set. 

In this letter we study two different simple  mono atomic glasses: 
a Lennard-Jones glass  (LJ) and 
a glass defined by an IC potential,  which  
is a modified LJ potential, designed to favor icosahedral
 local order \cite{Dzu}.
The IC glass has more pronounced structural ordering
than the LJ glass \cite{Dzu}. 
Furthermore, while  the LJ glass does not have a 
well defined glass transition, it is well defined  for the IC
glass \cite{Dzu1}. These differences leads one to expect differences in the 
organization of the stress network. 

We prepared systems of 21952 atoms  using molecular dynamics 
 simulations for both glasses.
 Periodic  boundary conditions were assumed. 
 Both potentials were studied extensively by numerical simulations.
 Details about systems' parameters and  preparation are given  in \cite{Kst} 
 for LJ. The parameters of IC potential are given  in \cite{Dzu} with density as in\cite{Dzu1}. 
 The initial LJ liquids were prepared and equilibrated at $T=1$. LJ glass was
 prepared by steepest  decent quench to zero temperature glass. 
  The initial IC liquids were prepared 
 and equilibrated at $T=1.6$. Glasses at various temperatures were prepared by coupling to a heat
  bath and then  equilibrating  the systems till the systematic decrease in energy stops. To subtract
  the effects of thermal noise, all the glasses where quenched to zero temperature by steepest
  descent procedure. We used 4 independent systems of IC glass and 3 independent systems of LJ glass.
  Results reported below are similar for all systems.
All the quantities are expressed in the LJ reduced units and the distances 
are given in atomic distances.

First, we briefly discuss the density correlation function
 $\rho(r)= n(r)/4 \pi r^2$.
 It is well known that in simple liquids,  long range decay of
  $\rho(r)$ is (e.g. \cite{Hen}) 
\begin{equation}\label{gr}
\rho(r)-\rho_0\sim \exp^{-\alpha_0 r}\cos(\alpha_1 r +\phi)
\end{equation} 
where  $\rho_0$ is the bulk value, $\alpha_0$ is decay length and $\alpha_1$ is the  wavelength of the
 decay. For simple liquids, $\alpha_0$   depends on the temperature. 

We calculated numerically  the variations of $\rho(r)$ for both potentials in the 
liquid phase and in zero temperature glass. In liquid phase, after the first coordination shell, 
$\rho(r)$ decays in accord to
eq. \ref{gr}.
Fig. 1 shows the variations in $\log(\rho(r)-\rho_0)$ for both potentials in the zero temperature  glass. 
After a few coordination shells,
the  glassy correlations exhibit decays
 of the type of eq. \ref{gr} with
$\alpha_1 \approx 1.12 \cdot 2  \pi,\alpha_0 \approx 0.56$ for IC glass and 
$\alpha_1 \approx 1.14 \cdot 2 \: \pi ,\alpha_0 \approx 0.62 $    for LJ glass at  $T=0$. 
A comparison with the liquids  at $ T = 1$ shows that the differences in the
 wavelength  $\alpha_1$
 are very minor, and  the variations in  the decay length $\alpha_0$ are about $20 \%$.
 Thus it is apparent that 
density fluctuations  at longer scales in glasses  are
similar to those in liquids. 
This result is in accord with the view that 'inherent' glasses and
liquids have a very similar structure but the temperature disguises it.
The main qualitative effect of the temperature  on $\rho(r)$ is
 to smooth structural details in the  2-4 coordination shells. 
 For similar results on different systems  see in\cite{Don}.

\begin{figure}
\bbox {(a)\epsfxsize=3.5truecm\epsfbox{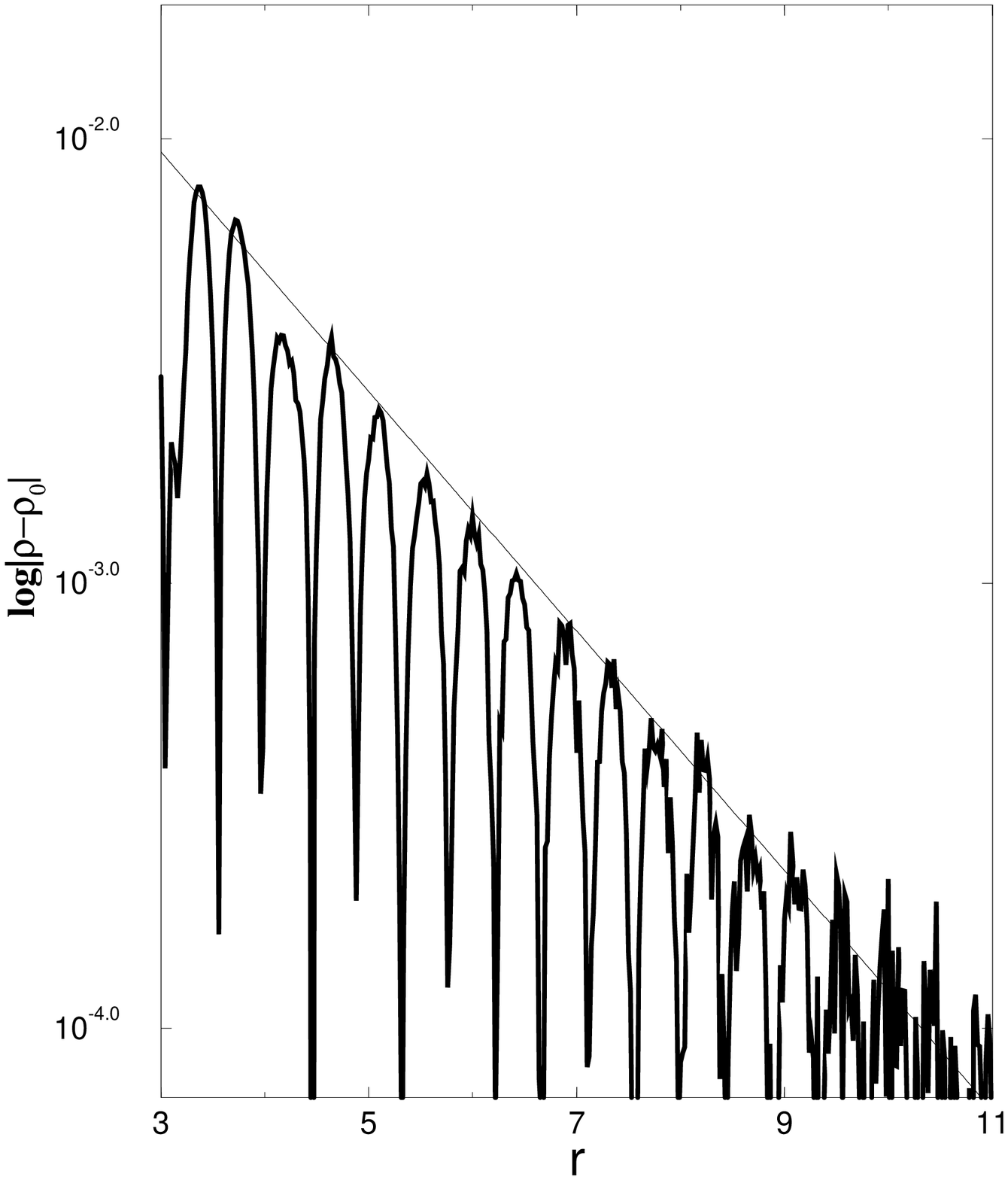}
(b)\epsfxsize=3.5truecm\epsfbox{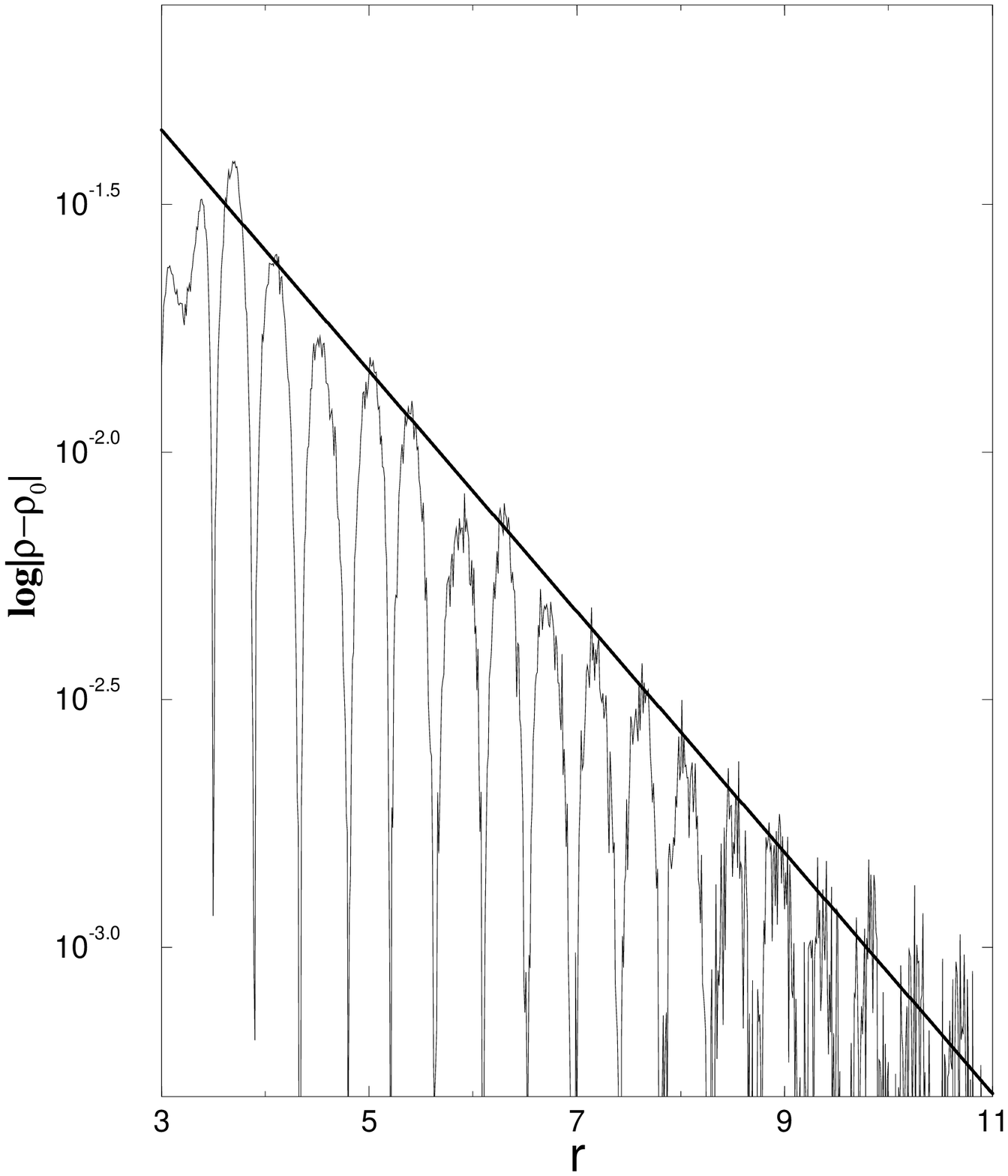}}
\vskip 0.2cm

\caption{ $\log(\rho(r)-\rho_0)$ vs $r$ for an  LJ (a) and  IC (b) 
glasses at 0k. We fit the decay exponents by the straight lines.  }
\end{figure}

As stated before, the probability distribution of $J$ is quite wide. Its width and 
 shape are similar for the LJ and the IC glasses.
To evaluate the effects of the stresses   
we calculated $n_J(r)$ for the extreme 
parts of the J distributions (4 percent of the particles).
As shown in Fig. 2, the  first 3 peaks of the conditional distributions
are  considerably narrower than the peaks of $n(r)$. The  peaks are shifted
relative to each other by about $0.1$ atomic distance.
This shift in the position is a result of the different compression of
 the environments. 
The shift toward smaller $r$ for small $J$'s 
indicates  compressed local environments whereas the shift 
toward larger $r$'s for large $J$'s
 indicates stretched local environments. The radiuses of those environments are consistent with
 effective potential estimates \cite{Kst}.
The shifts in $r$ in the first coordination shell are 
 analog to variations in volume in a Voronoi scheme.

The apparent differences between  LJ and IC glasses (Fig. 2a and 2b) 
are a manifestation of the 
different finer structural details. For the LJ glass,  the $n_J(r)$  
look similar but shifted regardless of the values of $J$, 
  indicating  the absence of a well defined local structure.  
However, in the IC glass case,  the width of the first peak is much
narrower for small $J$s than for the larger ones (a factor of about 2). 
The structural origin of this becomes even
 more apparent when the ratios between the positions of the
 $1^{\mbox{st}}$ and  $2^{\mbox{nd}}$ peaks
 in $n_J(r)$ are calculated. 
In the LJ glass the ratio is $\approx 1.7$ for both 
 high/small values of $J$.
 In the IC glass, the ratio is $ \approx 1.7$ for small $J$s and 
$ \approx 1.63$
 for large $J$'s.
 The ratio of 1.7 does not suggest any specific local structure,
 although it is very
 pronounced in IC glass, where  atoms barely appear in the interval
 between the peaks. 
 The 1.63 ratio corresponds to the distances between atoms on the faces of an
 Icosahedron! Thus 
the non-compressed environments are consistent with atoms that sit
 on the faces of empty icosahedra. 
Those icosahedra are the voids with  radius of one atomic distance
 observed in \cite{SDE}! On the other hand,
 the compressed environments have an almost full icosahedral nature, i.e. 
there are exactly 12 neighbors to an atom, and almost all 
atoms in the first neighbor 
 shell have five nearby neighbors.  The neighbors in the second peak 
of the compressed environments, 
are very sharply defined (the ratio between the stressed and compressed  $2^{\mbox{nd}}$ peaks
 is about 3.)  They are held by
the more diffuse, stretched empty icosahedral packing.
\begin{center}
\begin{figure}
\vspace{-0.6cm}
 (a)\epsfysize=5.5cm\epsfbox{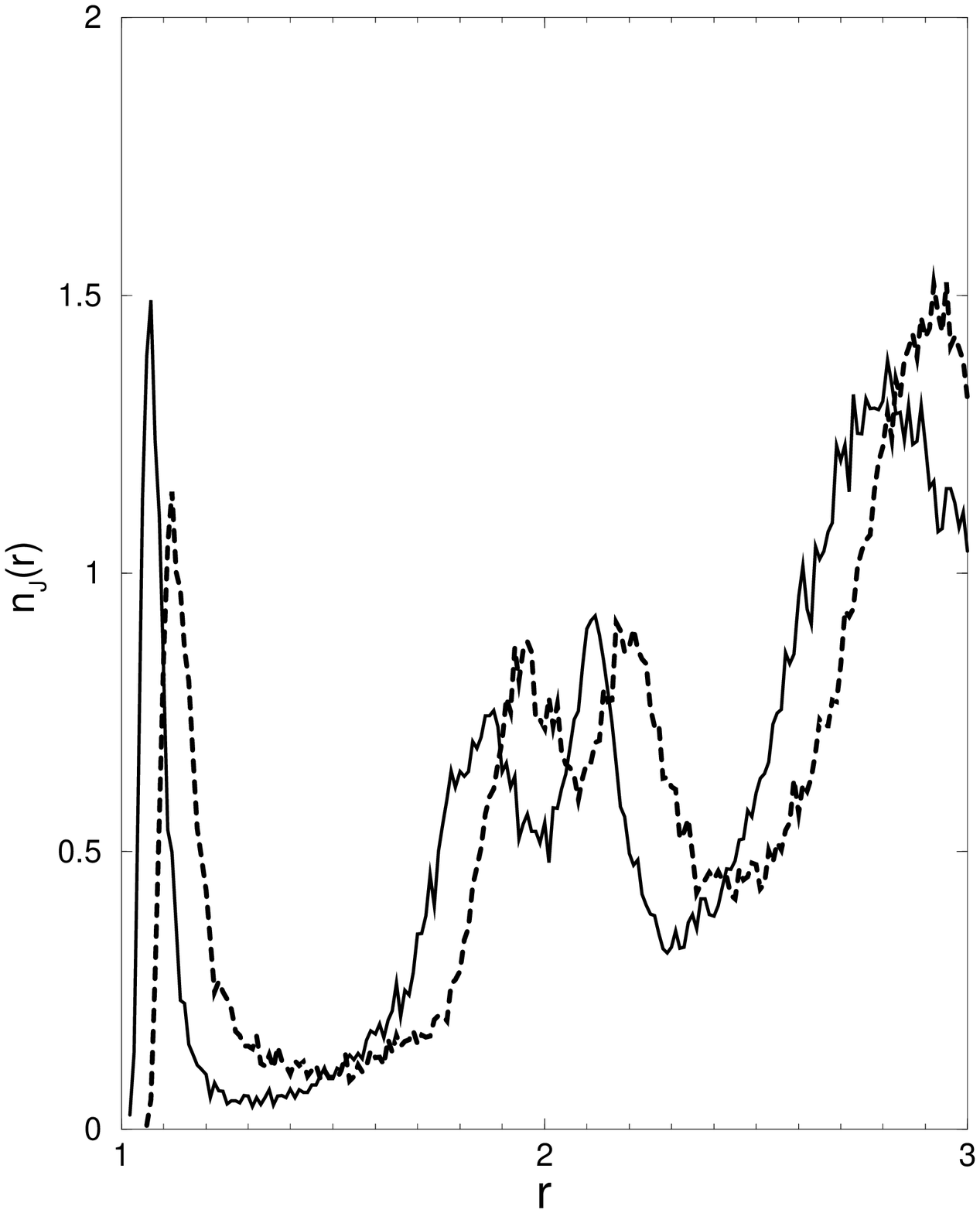}\\
 \vspace{-0.1cm}
(b)\epsfysize=5.5cm\epsfbox{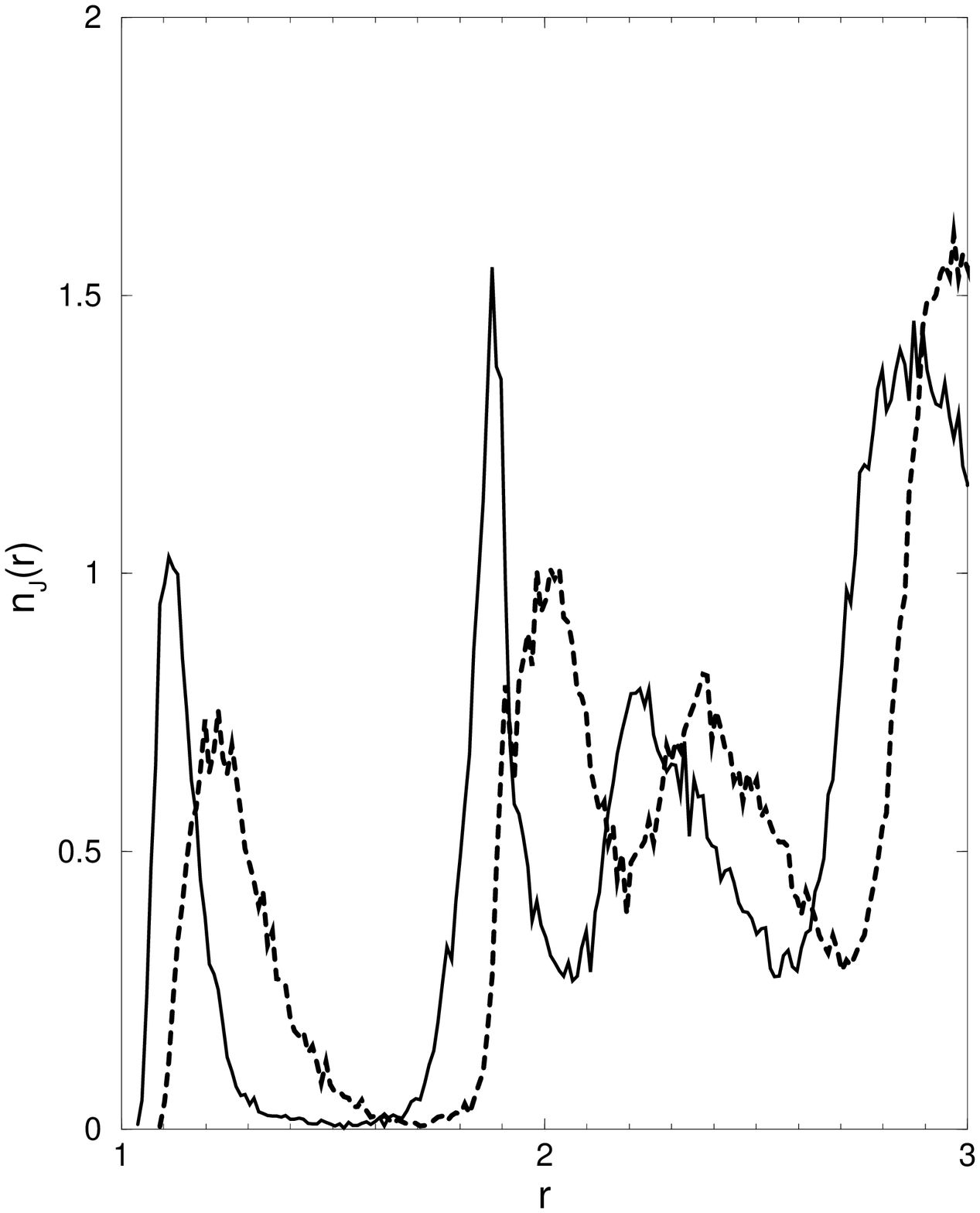}
%\vskip 0.2cm
\caption{ $n_J(r)$ for LJ  (a) and IC (b) glasses in the
first 3 coordination shells 
for the two extreme parts of the distribution of $J$'s. The solid line is
 for the low $J$s (compressed environments) and the dashed line is for
the high $J$s.}
\end{figure}
\end{center}
The stretched and compressed environments are spread throughout the system. 
Since the average number of neighbors in the glass is about 12,
the two sets
and their first neighbors span the whole glass.
Thus the two types of extremely stressed environments
define all of the glass. Note  
that in both glasses, the average nearest neighbor distances change as
a function of the distance from these sets of atoms. 
Those changes happen up to second/third  peak position in order  to balance
compressed environments with stretched ones and vice versa.
This is a result of radial gradients of the stress around these 
atoms. The tensorial nature of the glasses is
 manifested in the adjustment of
the local distances.  

The next obvious question is whether these special environments 
are correlated at a longer range.  
To answer this question, we use the functions $dn_J(r)$, 
defined in eq. \ref{dg}
 for extreme values of $J$, 
to identify variations from the 'average structure'.
%For a random choice of atoms (with set size $4\%$ of the atoms in the system)
%we could not observe correlations in $dn_J(r)$ neither in LJ glass nor in IC glass.
In Figure 3 we present  $dn_J(r)$
for the compressed sites of the LJ and the IC glasses.
 This difference indicates the existence 
of excess correlations between the subset
$J$ and the total lattice.
 Similar results have been  obtained for the stretched
subsets except for a phase shift in the position of the peaks.  
We checked that for randomly chosen sets of atoms 
one observes no correlations at all in $dn_J$, as is indeed expected.
The intensity of the difference in short distances
 is larger by a
factor of two in the IC glass.  
The correlation lengths can be seen from the integrals of $dn_J$.
In the IC glass they are constant up to half of the system size 
(14 atomic units). 
On the 
the LJ case the correlations decay from a scale of 2-3 until they are 
randomized  at a distance of 10.

\begin{center}
\begin{figure}[t]
\vspace{-0.6cm}
(a)
\epsfxsize=4.2cm\epsfbox{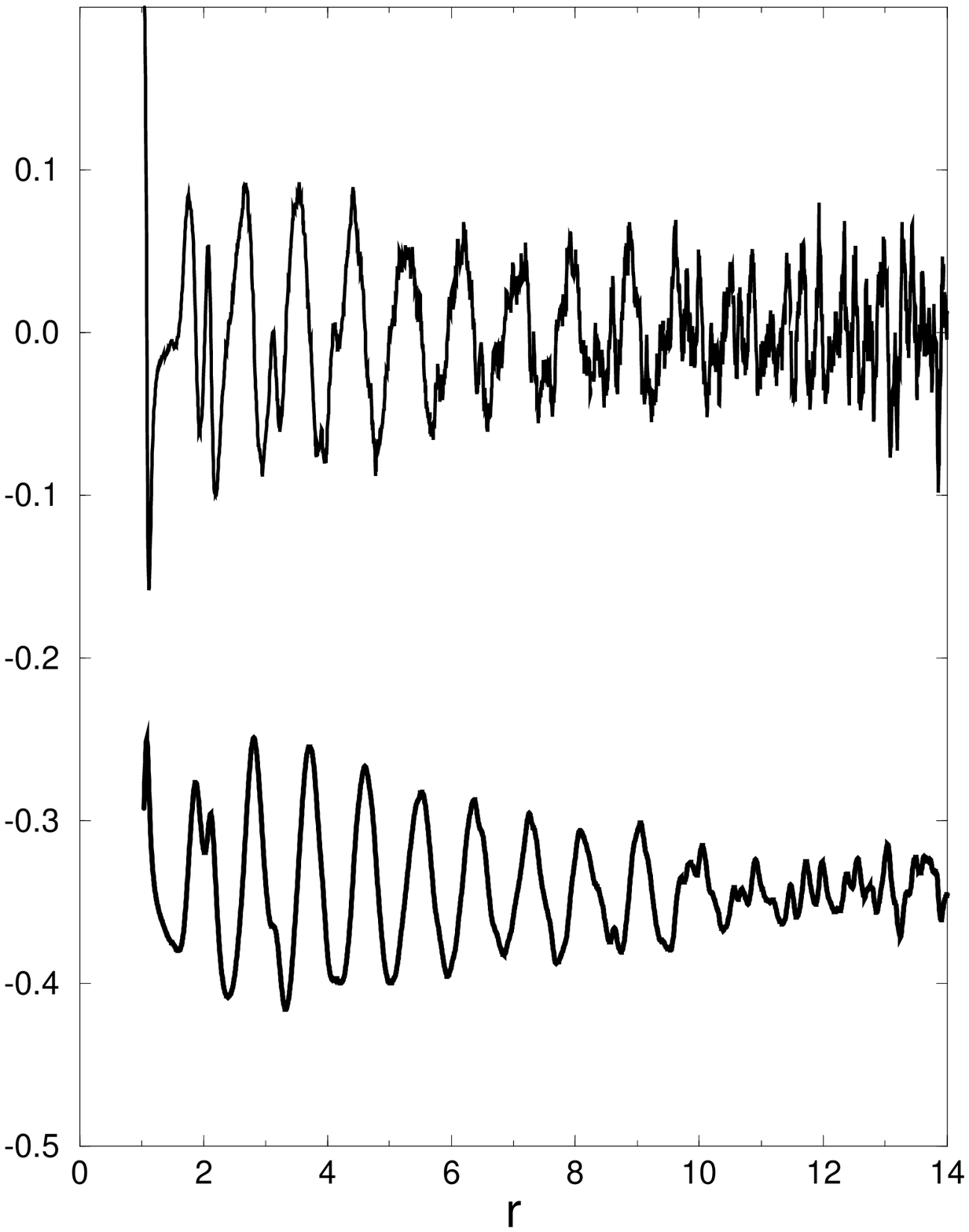}\\
\vspace{-0.1cm}
(b)
\epsfxsize=4.2cm\epsfbox{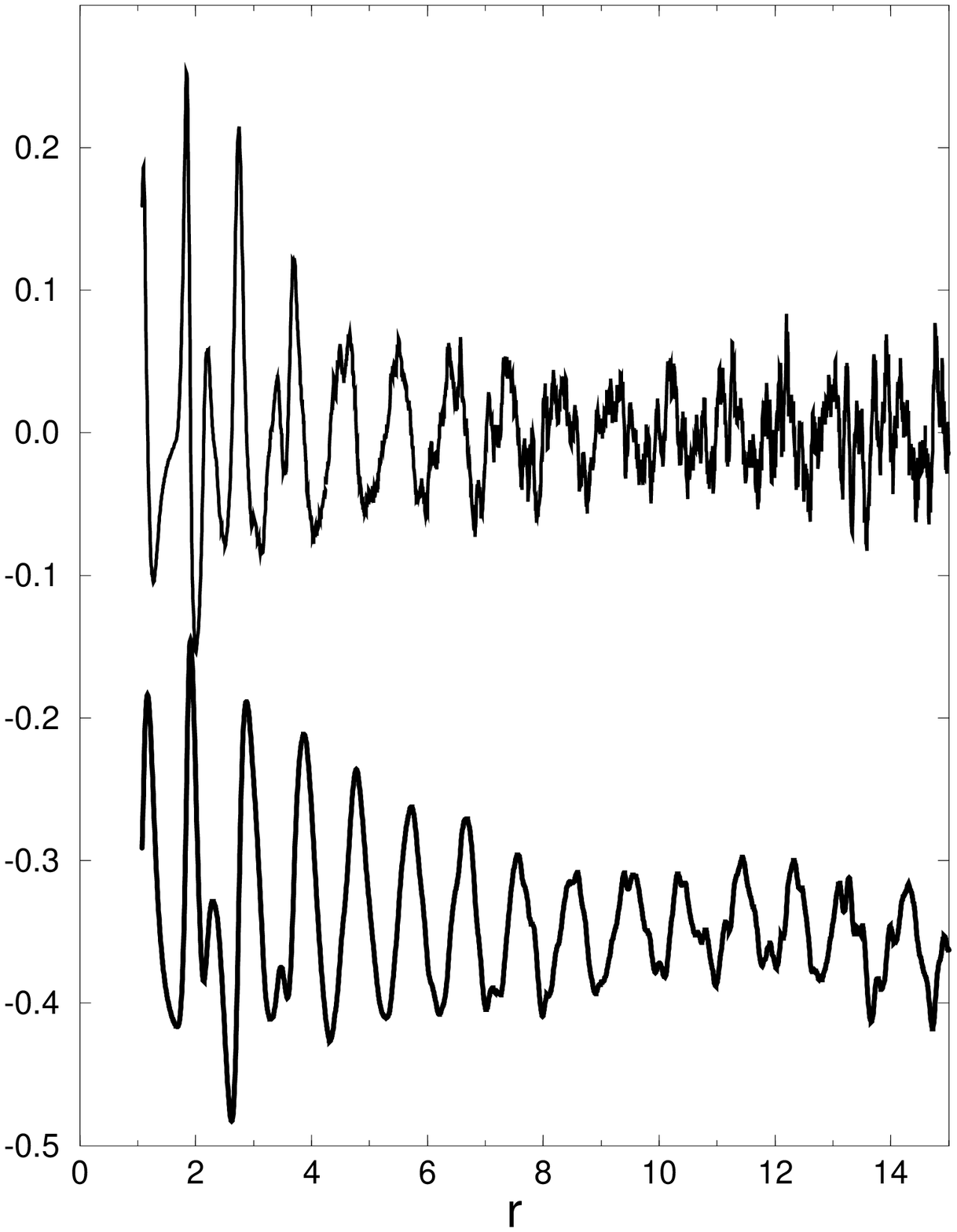}
\caption{$dn_J(r)$(upper curve) and its integral (shifted and in arbitrary 
scale) for
 the LJ  (a) and for the IC (b) glaseses. 
} 
\end{figure}
\end{center} 
We have already noted that while in the IC glass there is a 
distinct midrange structure,
it is absent in the LJ glass. 
This is the probable cause for the cutoffs in the correlations
in the LJ glass.
 The  limited range of excess correlations in the LJ glass is
also consistent with the lack  of glass transition in the 
LJ glass. 

The additional structure and additional correlations can also be observed in
the Fourier transform which also reduce the
statistical noise.  We performed Fourier transforms on the sets using
$ dn_J(q) = \sum _j \exp(i q r_j) dn_J(r_j) $.
In Figure 4
we present $|dn_J(q)|^2$ for lowest values of $J$ for LJ  and IC  glasses.
Very similar figures
were obtained for $|n_J(q)|^2$ for highest values of $J$.
A sharp pronounced peak at $q=q_{\mbox{peak}}$
exists is both glasses indicating that the period of the oscillations is about 
one atomic distance.
In the IC glass, $|n_J(q)|^2$ exhibits additional structure at
$q>q_{\mbox{peak}}$ which is absent in the LJ glass. The intensity of the peaks 
in the IC glass is much stronger. Inverse transforms confirmed the previous ranges
 of the
correlations.

\begin{figure}[t]
\bbox {(a)\epsfxsize=3.4truecm\epsfbox{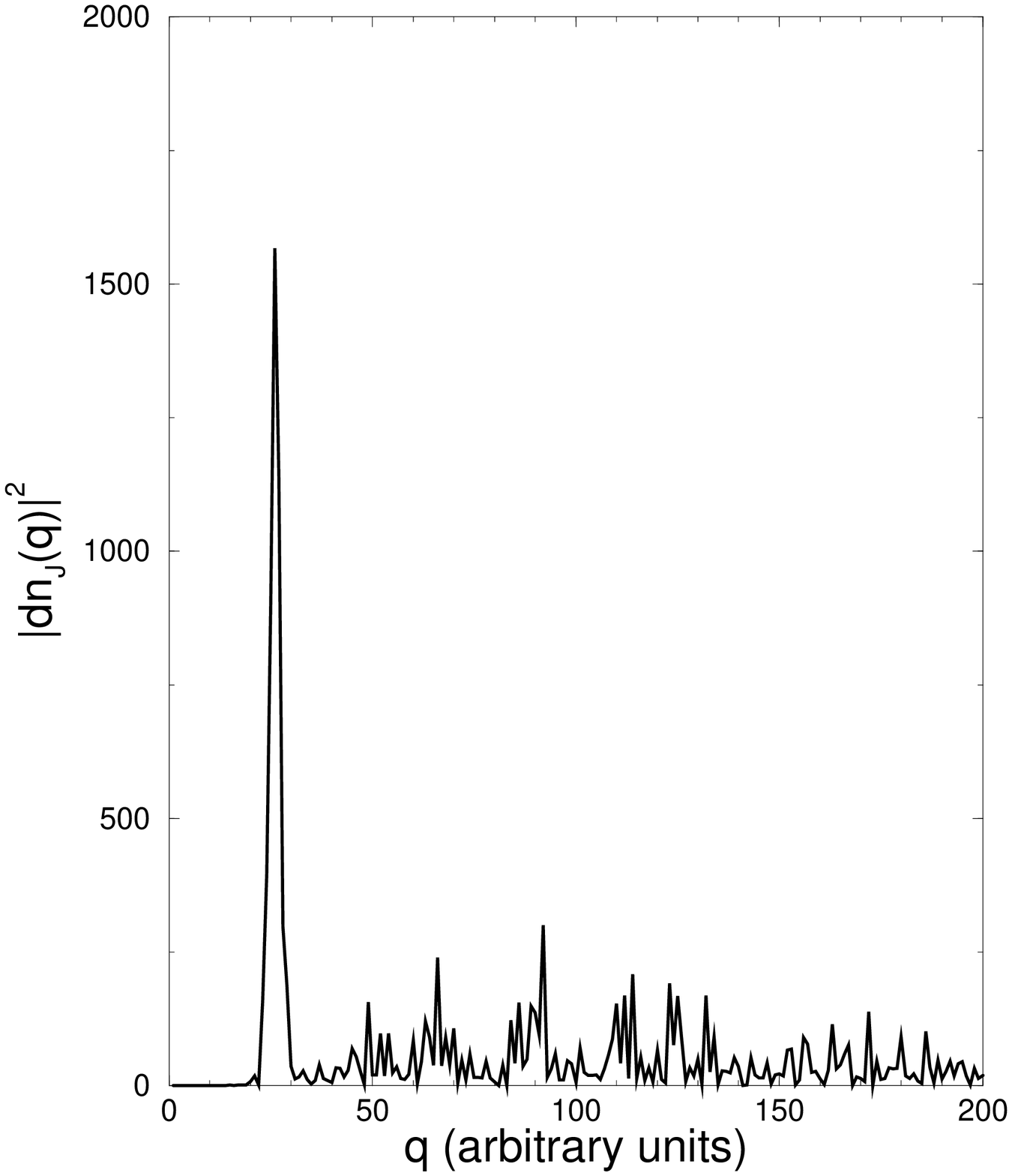}
(b)\epsfxsize=3.4truecm\epsfbox{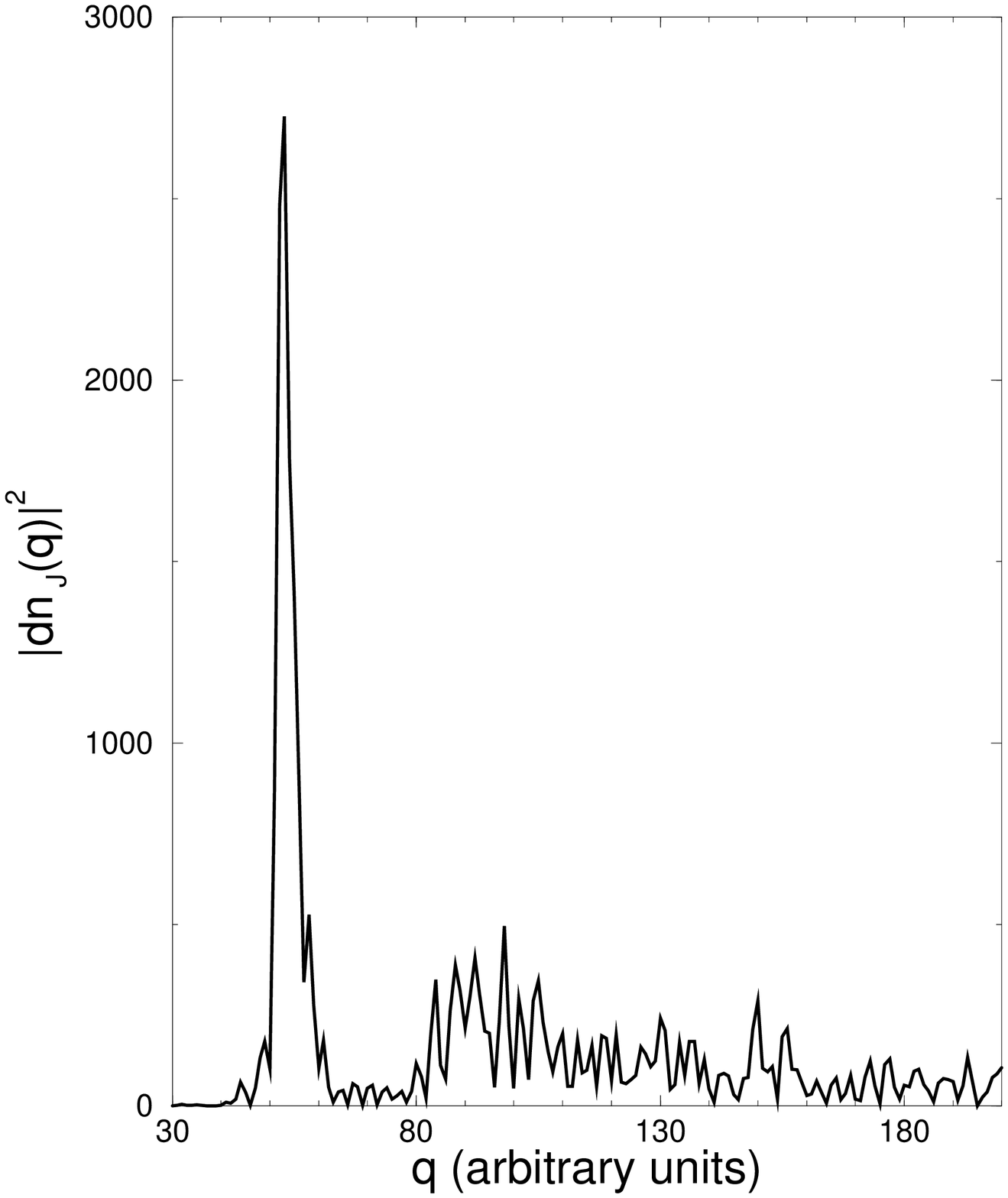}}
\vskip 0.2cm
\caption{$dn_J(q)^2$ for
 the LJ (a) and for the IC (b) glasses. } 
\end{figure}

The normalized conditional densities $dn_J(r)/(4\pi r^2)$
decay with the radius
as $r^{-2}$. The interpretation of these results is  
that though the total correlations in the system are randomized  
there are long range correlations within
the stressed networks. A possible explanation for such correlations
is the existence of correlated chains. 
 
In this letter we demonstrated how additional structural features, which are not apparent in
 $n(r)$ due to its wide peaks, emerge when one considers $n_J(r)$.
 For two simple glasses with very different short range
 structure, we have shown that whereas the averaged structural correlations
 have exponentially dumped decay analog to dense liquids,
 stronger and extended structural correlations are revealed
 throughout the system when one considers finer measures using LEP.
 
We have demonstrated the existence of two critical subsets of atoms in the
 structure.
 These subsets are not
random structures but a result of interrelations between local structure
(and packing) and
 local elastic features (see also ref.\cite{Kst}).
 We have shown that those sets induce long range
 structural correlations which  decay much slowly than the total 
 radial distribution function.
 These finding suggest a fascinating picture of simple glasses.

There are  natural extensions of this study. First, using LEP one 
can proceed to study more complicated
glasses, where the need to identify structural correlations
 is even more acute.
One might expect that for stronger type of local organization 
 there might be stronger and longer  range correlations. It
  is thus interesting to study different kind of 
potentials to see what are the universality classes 
of the decay of the conditional correlations.

It is natural to assume that atomic diffusion near the glass transition
 will be dominated by movements in the stretched environments. This is 
obviously related to the existence of  free volume due to the icosahedral 
voids\cite{Dzu2}. The existence of correlations between these states 
suggest that atomic diffusion, in temperatures  slightly 
above the glass transition, will occur as a correlated 
set of jumps in the  stressy network. Below the glass
transition such correlated movements are inhibited. 
Since one  expect  LEPs to vary smoothly at the glass transition,  this parameter
  is especially convenient for study and comparison of structural correlations in 
  glasses in a wide range of temperatures
as well as supper cooled liquids near glass  transition.

\subsection*{Acknowledgments}
We thank M. Dzugatov for providing information on the IC potential.
 We also thank S. R. Elliott and S. Safran  for their comments.


\begin{thebibliography}{9}
\bibitem{eli}
Uhlherr A, Elliott S.R., 1995  Philos. Mag. B {\bf 71}  611.
Uhlherr A , Elliott S.R. J Non-Cryst Solids {\bf 193} 98.
\bibitem{Don}
Donati C., Glotzer S. C. and Poole P. H., 1999, Phys. Rev. Lett. {\bf 82} 5064.
Donati C., Glotzer S. C., Poole P. H. Kob W. and Plimpton S. J., 1999, Phys. Rev. E. {\bf 60} 
 3107.
 \bibitem{Han}
J. P. Hansen and I. R. McDonald, {\em Theory of simple liquids}, 1986, AcademicPress, second
edition.
\bibitem{egami} 
Egami T., Madea K. and Vitek V., Phil. Mag. A. , {\bf 6}, 883 (1980).
 Egami  T, and Vitek V.,Phys. Stat. Sol., {\bf 144}, 145 (1987).
\bibitem{Alexander}
Alexander S., 1998, Phys. Rep. {\bf 296} 65.
\bibitem{Kst}
Kustanovich T. and Olami Z., 2000, PRB {\bf 61} 4813.
\bibitem{Kust1}
Kustanovich T., Alexander S. and  Olami Z., 1999, Physica A {\bf 266} 434.
\bibitem{Dzu}
Dzugutov M., 1992, Phys. Rev. A {\bf 46} R2984.
\bibitem{Dzu1}
Dzugutov M., 1993, Phys. Rev. Lett. {\bf 70} 2924.
\bibitem{Hen}
Henderson J. R. and Sabeur Z. A., 1992,  J. Chem. Phys. {\bf 97} 6750 and ref. there.
\bibitem{SDE}
Sadigh B., Dzugutov M., Elliott S. R., 1999, Phys Rev. B {\bf 59} 1.
\bibitem{Dzu2}
Dzugutov M., 1994, Europhys. Lett. {\bf 26} 533.

\end{thebibliography}
\end{document}